\documentclass{elsart}
\usepackage{graphicx}
\usepackage{epsfig}
\usepackage{xspace}
\usepackage{amssymb}
\usepackage{amsmath}
\begin{document}
\newlength{\caheight}
\setlength{\caheight}{12pt}
\multiply\caheight by 2
\newlength{\secondpar}
\setlength{\secondpar}{\hsize}
\divide\secondpar by 4
\newlength{\firstpar}
\setlength{\firstpar}{\secondpar}
\multiply\firstpar by 2
\def\bce{\begin{center}}  \def\ece{\end{center}}
\def\bit{\begin{itemize}}    \def\eit{\end{itemize}}
\def\ben{\begin{enumerate}}    \def\een{\end{enumerate}}
\newcommand{\hepph}[1]{{\tt hep-ph/#1}}
\newcommand\cpc[3]{{\it Comput. Phys. Commun. }{\bf #1} (#2) #3}
\newcommand\npb[3]{{\it Nucl. Phys. }{\bf B #1} (#2) #3}
\newcommand\annp[3]{{\it Ann. Phys. }{\bf #1} (#2) #3}
\newcommand\plb[3]{{\it Phys. Lett. }{\bf B #1} (#2) #3}
\newcommand\prd[3]{{\it Phys. Rev. }{\bf D #1} (#2) #3}
\newcommand\sjnp[3]{{\it Sov. J. Nucl. Phys. }{\bf #1} (#2) #3}
\newcommand\jetp[3]{{\it Sov. Phys. JETP }{\bf #1} (#2) #3}
\newcommand\zpc[3]{{\it Z. Physik }{\bf C #1} (#2) #3}
\newcommand\ibid[3]{{\it ibid. }{\bf #1} (#2) #3}
\newcommand\jhep[3]{{\it JHEP }{\bf #1} (#2) #3}
\newcommand\ptp[3]{{\it Prog. Theor. Phys. }{\bf #1} (#2) #3}
\newcommand\cal{\mathcal}
\hyphenation{BA-BA-YA-GA}
\hfill
\parbox[0pt][\caheight][t]{\secondpar}{\rightline{\tt \shortstack[l]
{FNT/T 2001/06}}}  
\begin{frontmatter}
\title{An improved Parton Shower algorithm in QED}
\author[add1,add2]{Carlo Michel Carloni Calame}
\address[add1]{Dipartimento di Fisica Nucleare e Teorica, Universit\`a di Pavia}
\address[add2]{INFN, Sezione di Pavia}
\centerline{\small Via A. Bassi, 6 \ \  - \ \  27100 Pavia, Italy}
\begin{abstract}
An improved QED Parton Shower algorithm to calculate photonic radiative
corrections to QED processes at flavour factories is described.
We consider the possibility of performing photon generation in
order to take into account also the effects due to interference between initial
and final state radiation. Comparisons with exact order $\alpha$ results are
shown and commented.
\end{abstract}
\begin{keyword}
Parton Shower, YFS, Event Generation
\end{keyword}
\end{frontmatter}
\section{Introduction}
\label{intro}
The precise luminosity monitoring and data analysis at $e^+e^-$ flavour 
factories (DA$\Phi$NE, PEP-II, KEKB, CESR) require that cross section
calculations for QED final state processes 
($e^+e^-\to e^+e^-$, $\mu^+\mu^-$, $\gamma\gamma$) have theoretical error
below the 1\% level. Also Monte Carlo simulations with the same accuracy are
strongly demanded for data analysis. This implies to include the relevant
radiative corrections, in particular the large effects due to multiple photon
emission.

Aiming at providing a useful tool for data analysis, a Monte Carlo event
generator (BABAYAGA) for QED processes at flavour factories has been
developed. An exhaustive description of its main features and its basic
theoretical ingredient, the QED Parton Shower (PS), can
be found in \cite{ournpb}. Here, we remind only of the basics of the
theoretical approach which the generator is based on, in order to describe
recent improvements introduced in the code to give a more precise
simulation also for exclusive radiative events (i.e. with observed photons in
the final state). They concern the inclusion of
the effects due to interference between initial state and final state radiation,
which are not present in a pure PS framework.

The outline of the paper is the following: in the next section, the main 
theoretical framework is summarized and the Parton Shower in QED is presented.
In section \ref{phkine}, the generation of photons kinematics is revised and a
new generation, inspired to the Yennie-Frautchi-Suura formula, is introduced. In
section \ref{CandR}, the results obtained by the PS are compared to known
perturbative results and the improvements due to the new generation are shown.
The accuracy of the approach is also estimated. In the concluding section, the
present study is summarized and the open issues and possible developments are
discussed.
%

\section{The Parton Shower algorithm in QED}
\label{QED-PS}

A widely used theoretical tool, to take into account the bulk 
of the photonic radiative corrections, consists in the calculation of the
corrected cross section following the master formula \cite{prd}
\begin{eqnarray}
\sigma(s)=&\int dx_- dx_+ dy_- dy_+ \int d\Omega_{cm}
D(x_-,Q^2)D(x_+,Q^2) \times & \nonumber\\ 
&D(y_-,Q^2)D(y_+,Q^2) \frac{d\sigma_0}{d\Omega_{cm}}
\big(x_-x_+s,\vartheta_{cm}\big)\Theta(cuts) \ \ , & 
\label{eq:sezfs}
\end{eqnarray}
which is based on the factorization theorems of universal infrared and collinear
singularities. In the previous formula, $d\sigma_0/d\Omega$ represents the
Born-like differential cross section for the process under consideration and
$D(x,Q^2)$ are the electron structure functions (SF) for initial and final
state radiation.
Equation (\ref{eq:sezfs}) is also suited for the Monte Carlo generation of
unweighted events.

The SF's describe the effects of
multiple emission of soft and hard photons in the collinear limit to all orders
of perturbation theory. The SF is the solution of the QED
Dokshitzer-Gribov-Lipatov-Altarelli-Parisi (DGLAP) evolution equation \cite{ap}
in the non-singlet channel
\begin{equation}  
Q^2\frac{\partial}{\partial Q^2}D(x,Q^2)=
\frac{\alpha}{2\pi}\int_x^{1}\frac{dy}{y}P_+(y) D(\frac{x}{y},Q^2) \ \ ,  
\label{eq:ap} 
\end{equation}
where $P_+(x)$ is the regularized $e\to e+\gamma$ splitting function. 
The electron SF has a clear and intuitive meaning: it represents the 
probability density  of finding ``inside'' a parent electron an electron with
momentum fraction $x$ and virtuality $Q^2$.

In the literature (see for example \cite{rnc} and references therein) several
solutions of DGLAP equation are available, but we focus our attention on its
Monte Carlo solution, known as the Parton Shower algorithm
\cite{psqcd,psqed,psohl}. The advantages of the PS are mainly two: it is an 
{\it exact} numerical solution of the DGLAP equation and it allows an 
approximate generation of the photons momenta. The steps of the algorithm can
be extracted introducing the Sudakov form factor \cite{sff} and, using it,
getting an  iterative solution of DGLAP equation. Namely, the solution is
\begin{eqnarray} 
D(x,Q^2)&=&\Pi(Q^2,m^2)\delta(1-x)+ \nonumber\\
&+& \int_{m^2}^{Q^2}\Pi (Q^2,s')\frac{d s'}{s'}\Pi (s',m^2)\frac{\alpha}{2\pi}
\int_0^{x_+} dy P(y) \delta (x-y)+ \nonumber\\
&+& \int_{m^2}^{Q^2}\Pi (Q^2,s')
\frac{ds'}{s'}\int_{m^2}^{s'}\Pi (s',s'')\frac{ds''}{s''}
\Pi (s'',m^2)\times \nonumber \\
&\bigg(&\frac{\alpha}{2\pi}\bigg)^2\int_0^{x_+} dx_1\int_0^{x_+}
dx_2 P(x_1)P(x_2) \delta (x-x_1x_2) + \bigg(\frac{\alpha}{2\pi}\bigg)^3\cdots    
\label{eq:alpha2}             
\end{eqnarray}
where 
$\Pi(s_1,s_2)=\exp [-\alpha/2\pi\int_{s_2}^{s_1}ds'/s'\int_0^{x_+}dzP(z)]$ is
the Sudakov form factor, which represents the probability that the electron
evolves from virtuality $s_2$ to virtuality $s_1$ with no emission of photons
of energy fraction greater than (an infrared regulator) $\epsilon = 1-x_+$.
Note that the eq. (\ref{eq:alpha2}) accounts for soft + virtual and real 
photons radiation up to all order of perturbation theory, in the leading
logarithmic approximation.
%
Concerning the soft + virtual cross section, it is worth noticing that, by 
setting the scale $Q^2$ to be equal to $st/u$, the Sudakov form factor 
exponentiates the leading logarithmic contribution of the $\cal{O}(\alpha)$
soft + virtual cross section 
as well as the dominant contribution coming from the infrared cancellation
between the virtual box and the initial-final state interference of the
bremsstrahlung diagrams. For a more detailed discussion on this topic, we
refer to \cite{ournpb}.

For the steps of our implementation of the PS algorithm, they can be found in
\cite{ournpb}. For the scope of this paper, it is sufficient to notice that
the algorithm simulates a shower of photons emitted by the electron following
the solution of eq. (\ref{eq:alpha2}). When the algorithm stops, we are left 
with the energy fraction $z_i$ of each photon (distributed according to the 
Altarelli-Parisi splitting function), the virtualities of the electron at each
branching and the remaining energy fraction $x$ of the electron after the 
showering. The $x$ variable is distributed according to $D(x,Q^2)$. By means of
these quantities, an approximate branching kinematics can be obtained, as
discussed below.
%

It is worth stressing that, when an event sample has been generated
following equations (\ref{eq:sezfs}) and (\ref{eq:alpha2}), the sharing of the
sample in elastic events (which correspond to soft + virtual cross section),
one-photon events, two-photons events and so on,
is automatic and built-in in the method itself. On the event sample, {\em any}
experimental cut can be imposed, by means of the $\Theta$ function which is
present in eq. (\ref{eq:sezfs}).

The reliability of the PS method to simulate and generate events can be checked
by a systematic comparison of its results with known perturbative results, in
particular with exact ${\cal O}(\alpha)$ matrix element (as discussed in section
\ref{CandR}). The comparison allows to keep under control the
approximations intrinsic into the PS approach and to estimate its theoretical
accuracy. In order to perform such tests,
an up to ${\cal O}(\alpha)$ PS algorithm has been developed as well. It allows
to calculate the corrected cross section of eq.~(\ref{eq:sezfs}) up to
${\cal O}(\alpha)$. Such a calculation is strongly required for fully
consistent comparisons between the PS predictions and the exact
${\cal O}(\alpha)$ ones.
%
%
The steps required for the 
${\cal O}(\alpha)$ PS, which are described in \cite{ournpb}, can be obtained by
using eq.~(\ref{eq:alpha2}) and expanding up to ${\cal O}(\alpha)$ the product 
$D(x_-,Q^2)D(x_+,Q^2)D(y_+,Q^2)D(y_-,Q^2)$ present in eq.~(\ref{eq:sezfs}).
%
%
\section{Branching kinematics}
\label{phkine}
The main advantage of the PS algorithm with respect to the collinear treatment
of the electron evolution is the possibility of going beyond the
strictly collinear approximation and generating transverse momentum $p_\perp$
of electrons and photons at each branching. In fact, the kinematics 
of the branching process $e(p) \to e'(p') + \gamma(q)$ can be written as
\begin{equation}  
p=(E, \vec{0}, p_z) \ \ , \ \  
p'=(zE,  \vec{p}_\perp,  p'_z) \ \ , \ \  
q=((1-z)E,  - \vec{p}_\perp,  q_z) \ \ .   
\label{eq:kinealt} 
\end{equation}
Once the variables $p^2$,  ${p'}^2$ and $z$ are generated by the PS algorithm, 
the on-shell condition $q^2=0$, together with the longitudinal momentum 
con\-ser\-va\-tion, al\-lows to obtain an expression for the $p_{\perp}$
variable:
\begin{equation}
p^{2}_{\perp}=(1-z)(zp^2-p^{\prime 2}) \ \ 
\label{eq:kinep}
\end{equation}
at first order in $p^2 / E^2 \ll 1$, $p^2_\perp / E^2 \ll 1$. From now on, we
will refer to eq. (\ref{eq:kinep}) as the ``PS prescription''.

Some 
%
not correct behaviours of the exclusive photon kinematics reconstruction
%
are connected with this PS picture,
%
due to the approximations intrinsic in eq. (\ref{eq:kinep}).
%
First of all, since
within the PS algorithm the generation of $p^{\prime 2}$ and $z$ are
independent, it can happen that in some branching the $p^2_{\perp}$ as given
by eq. (\ref{eq:kinep}) is negative. In order to avoid this problem, the
introduction of any kinematical cut on $p^2$ or $z$ generation (or the
regeneration of the whole event) would mean a not correct reconstruction of
the SF $x$ distribution, which is important for a precise cross section
calculation. Furthermore, in the PS scheme, each fermion produces its photons
cascade independently from the other ones, missing the effects due to the
interference of radiation coming from different charged particles. As far as
inclusive cross sections
%
(i.e., no cuts are imposed on the generated photons)
%
are considered, these effects are integrated out and
we do not expect them to be large, but when exclusive photon variables
distributions are looked at, they may be important.
 
Concerning the first problem, it can be overcome 
%
choosing a generation of photons $p_\perp$ different from eq. (\ref{eq:kinep}).
%
For example, we can choose to extract the photon 
$\cos\vartheta_\gamma$ according to the universal leading poles $1/p\cdot k$
present in the matrix element for photon emission (in the following referred to
as the ``LL prescription''), 
%
as done for example in \cite{psohl}. 
%
Namely, we can generate $\cos\vartheta_\gamma$ as
\begin{equation}
\cos\vartheta_\gamma \propto \frac{1}{1-\beta\cos\vartheta_\gamma} \ \ ,
\label{cthg_LL}
\end{equation} 
where $\beta$ is the speed of the emitting particle. In this way, photon
energy and angle are generated independently, differently from eq. 
(\ref{eq:kinep}). The nice feature of this prescription is that
$p^2_{\perp}=E^2_\gamma\sin^2\vartheta_\gamma$ is always well defined and the 
$x$ distribution reproduces exactly the SF, because we do not need to impose
further kinematical cuts to avoid unphysical events. At this stage, the
PS is used only to generate photons energies and photons multiplicity.
The problem of including the radiation interference is still unsolved, because
the variables of photons emitted by a fermion are still not correlated to the
other charged particles. The phenomenological comparison between PS and LL
prescriptions will be shown in the next section.

The issue of including photon interference can be successfully worked out 
looking at the Yennie-Frautschi-Suura (YFS) formula \cite{YFS,Muta}:
\begin{equation}
d\sigma_n\approx d\sigma_0\frac{e^{2n}}{n!}\prod_{l=1}^n\frac{d^{\it 3}
\mathbf{k}_l}{(2\pi)^32k^0_l}\sum_{i,j=1}^N\eta_i\eta_j\frac{-p_i\cdot p_j}
{(p_i\cdot k_l)(p_j\cdot k_l)} \ \ .
\label{YFS}
\end{equation}  
It is a very general formula which gives the differential cross section
$d\sigma_n$ for the emission of $n$ photons, whose momenta are 
$k_1,\cdots,k_n$, from a
kernel process described by $d\sigma_0$ and involving $N$ fermions, whose 
momenta are $p_1,\cdots,p_N$. In eq. (\ref{YFS}), $\eta_i$ is a charge
factor, which is $+1$ for incoming $e^-$ or outgoing $e^+$ and $-1$ for
incoming $e^+$ or outgoing $e^-$. Note that eq. (\ref{YFS}) is valid in the
soft limit ($k_i\to 0$). The important point is that it also accounts for
coherence effects. From YFS formula, it is straightforward to read out the
angular spectrum of the $l^{th}$ photon (``YFS prescription''):
\begin{equation}
\cos\vartheta_l\propto-\sum_{i,j=1}^N\eta_i\eta_j
\frac{1-\beta_i\beta_j\cos\vartheta_{ij}}{(1-\beta_i\cos\vartheta_{il})
(1-\beta_j\cos\vartheta_{jl})} \ \ .
\label{cthYFS}
\end{equation}

It is worth noticing that in the LL prescription, the same quantity writes as
\begin{equation}
\cos\vartheta_l\propto\sum_{i=1}^N\frac{1}{1-\beta_i\cos\vartheta_{il}}
\label{cthLL}
\end{equation}
whose terms are of course contained in eq. (\ref{cthYFS}).

In order to consider also coherence effects in the angular distribution of the
photons, we 
may generate $\cos\vartheta_\gamma$ according to eq. (\ref{cthYFS}), rather 
than eq. (\ref{cthLL}). We expect that the improvements on photon angular
distribution will be sizable, as discussed in the next section.

For the sake of clearness, it is worth stressing that we are not exploiting the
so called ``YFS exclusive exponentiation'' procedure; we are simply considering
the photons angular spectrum implied by equation (\ref{YFS}) and we are
generating the photons angular variables according to the distribution
(\ref{cthYFS}). Following such a procedure, we are embedding the more accurate
angular spectrum (\ref{cthYFS}) in the framework of the PS algorithm. If
the reader is interested in the different approach of the YFS exclusive
exponentiation, it is extensively and exhaustively treated in the papers of
reference \cite{jadach}.

\section{Comparisons and results}\label{CandR}

In this section, we deal with the application of the PS, mainly in its 
${\cal O}(\alpha)$ realization, to the Bhabha 
%
and radiative Bhabha 
%
process at
large angle, as implemented in BABAYAGA. To test the theoretical accuracy of 
the PS approach in total cross section calculation as well as in event
generation, an exact ${\cal O}(\alpha)$ calculation has been carried out in the
program LABSPV, used as a benchmark calculation. The code is described in
\cite{ournpb,sabspv} and it is based on the formulae given in \cite{exsv,BK}.

\begin{figure}
\bce
\epsfig{file=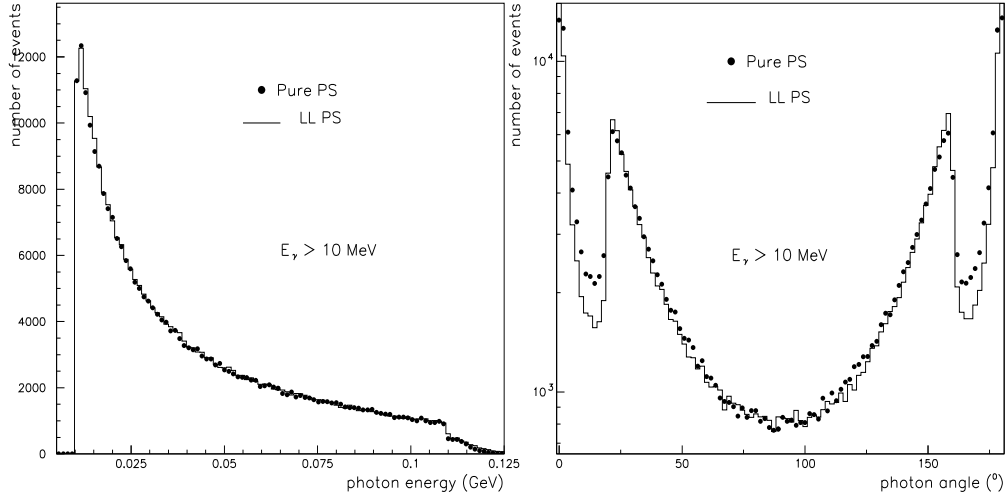,width=14cm,height=7.5cm}
\ece
\caption{Comparison of energetic (on the left) and angular (on the right) photon
spectrum given by ${\cal O}(\alpha)$ PS with PS and LL prescription for photon
$p_{\perp}$ generation.}
\label{fig:psvsll}
\end{figure}

As first comparison, we consider the results of PS and LL prescriptions
on photon
distributions. They are obtained by means of the PS algorithm truncated at
${\cal O}(\alpha)$. We generated a sample of radiative Bhabha events at 
a typical DA$\Phi$NE energy
($\sqrt{s}=M_\Phi$) and requiring the final state $e^+$ and $e^-$ to lie within 
$20^\circ$ and $160^\circ$, with at least $0.4$ GeV of energy and a maximum 
acollinearity between them of $10^\circ$. Furthermore, we require that the 
photon has at least $10$ MeV of energy. For those unphysical $p^2_\perp < 0$
events generated in the pure PS prescription, we arranged to set the photon
collinear to the
emitting particle ($p^2_\perp = 0$), without regenerating the event. In fig. 
\ref{fig:psvsll}, the energy and angular distributions of the photon are
plotted. We note that, except some details, the two angular generation
schemes give roughly the same answer: this is consistent with the PS picture,
based on the leading logarithmic treatment of the branching kinematics.

\begin{figure}
\bce
\vspace{7.5cm}
\includegraphics{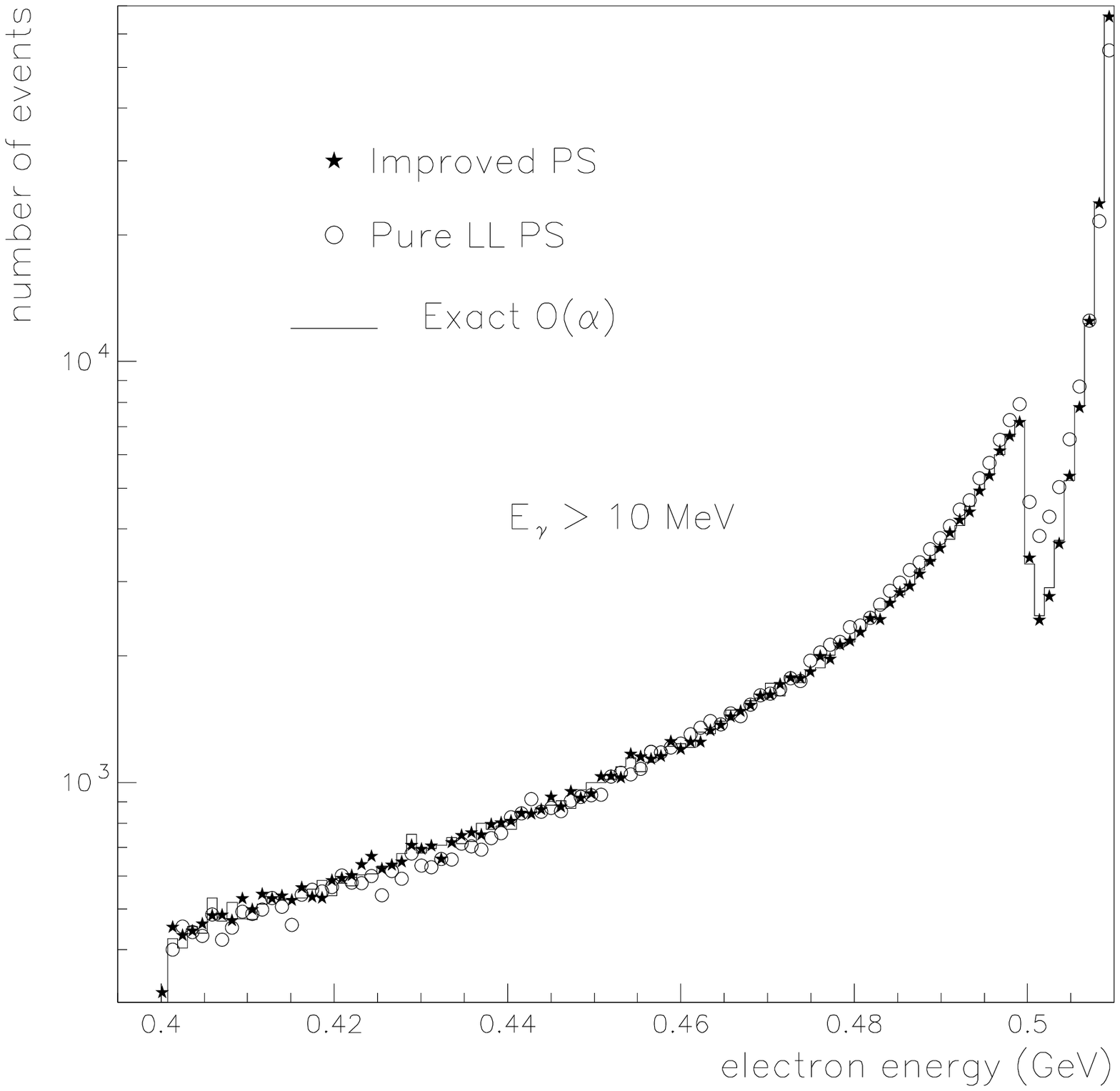}
\includegraphics{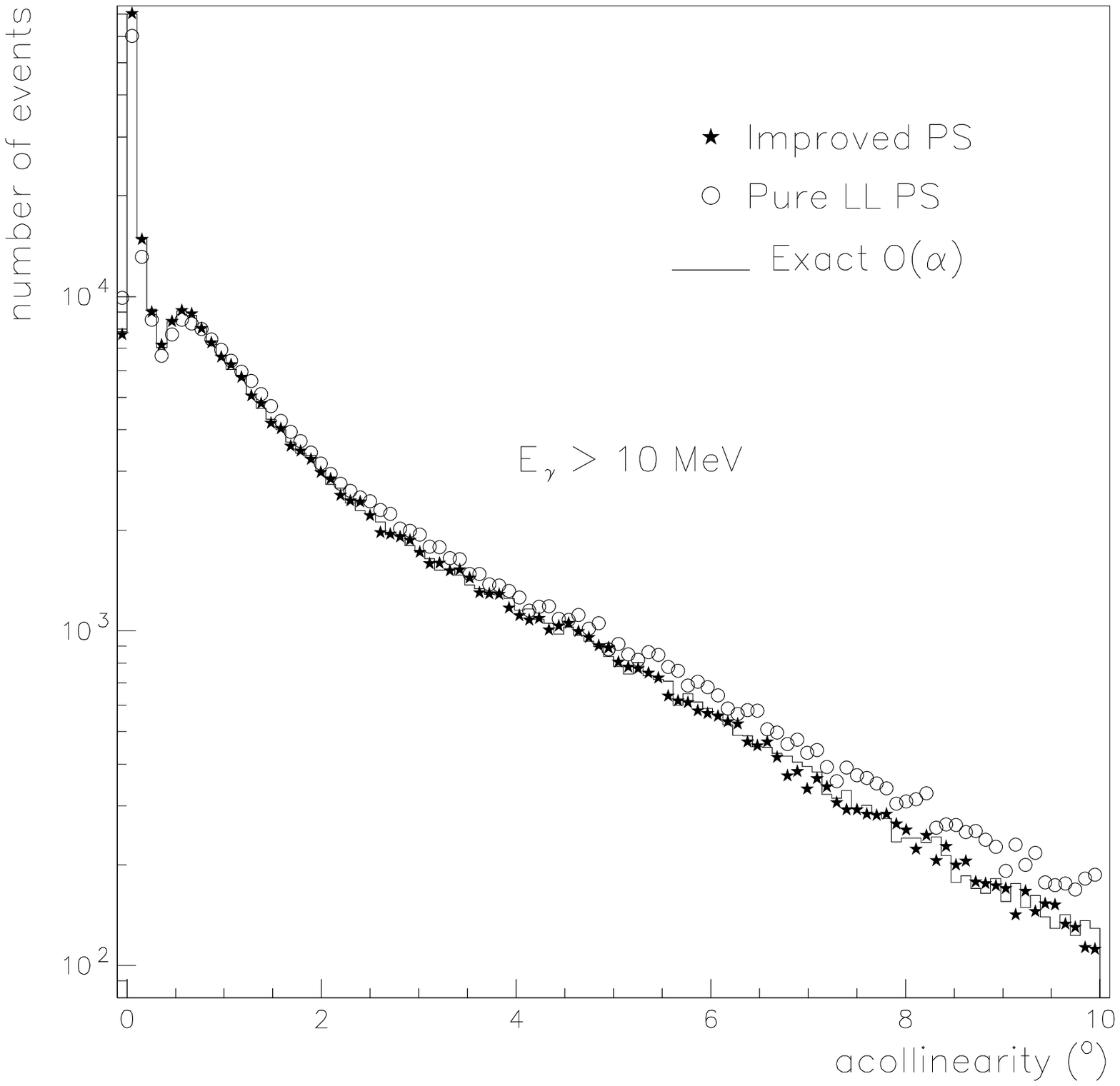}
\ece
\caption{Comparison of electron energy (on the left) and  acollinearity (on
the right) distributions obtained from LL and YFS prescriptions and exact 
${\cal O}(\alpha)$ matrix element.}
\label{fig:em_dist}
\end{figure}

Next, we consider how the distributions change if we weight photon angle by
eq. (\ref{cthYFS}) and we compare the results to the exact 
${\cal O}(\alpha)$ distributions.
\begin{figure}
\bce
\vspace{7.5cm}
\includegraphics{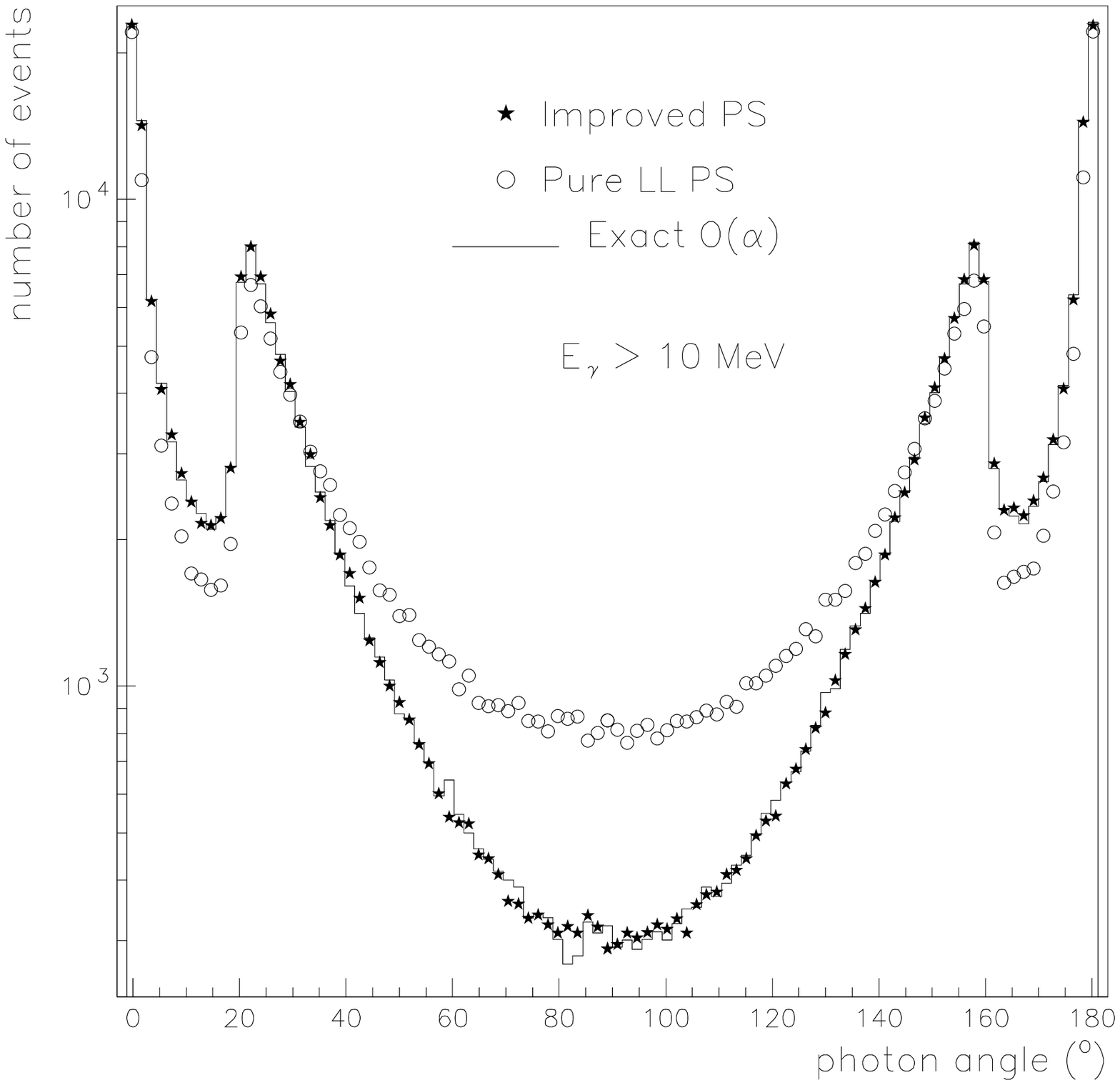}
\includegraphics{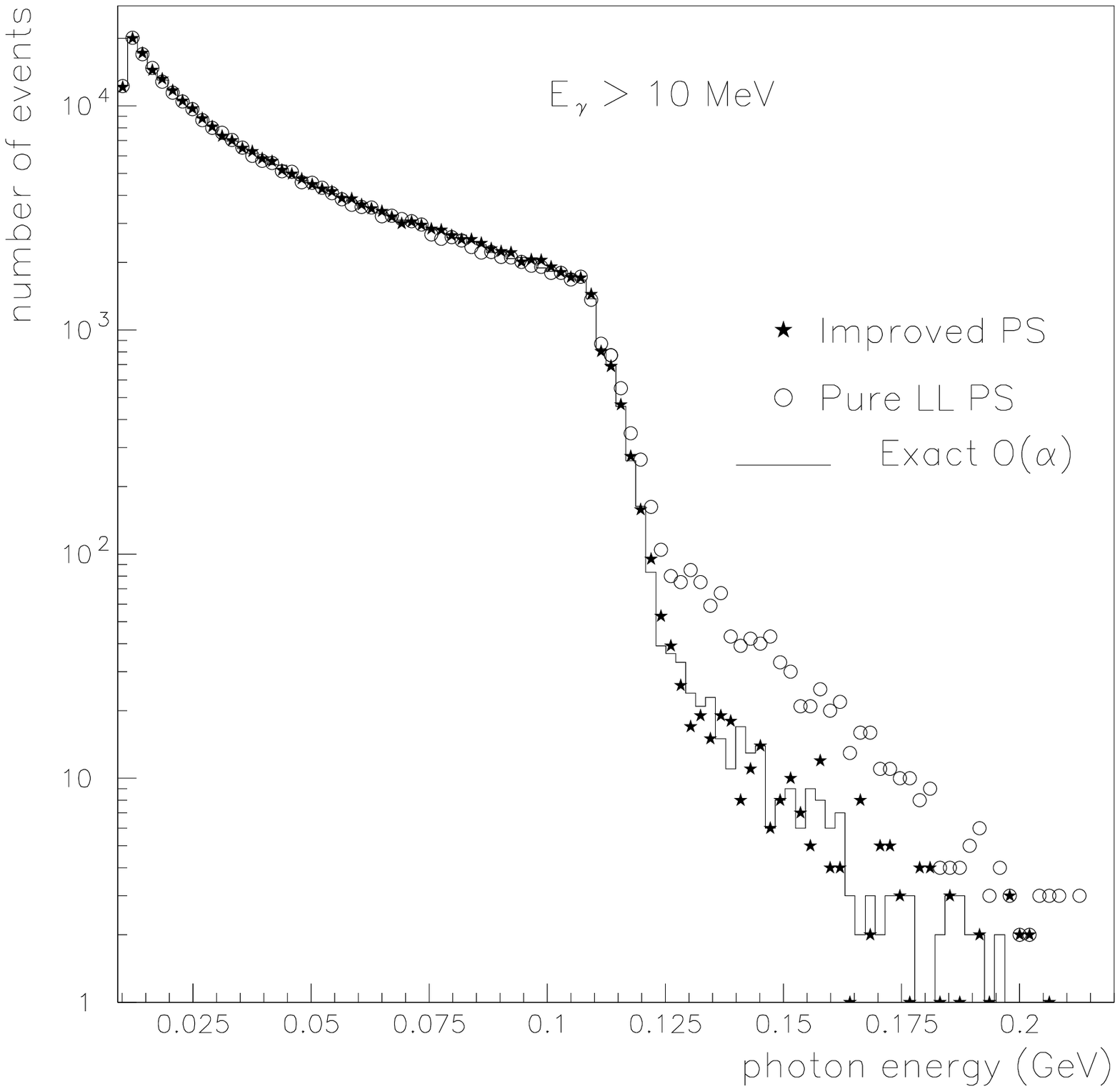}
\ece
\caption{Comparison of the angular (on the left) and energetic (on the right) 
photon spectra obtained from LL and YFS prescriptions and exact 
${\cal O}(\alpha)$ matrix element.}
\label{fig:exvspsvsll}
\end{figure}
On the left, fig. \ref{fig:exvspsvsll} shows the angular distribution of
the photon, imposing the same event selection criteria of the previous plots: 
the open circles represent the LL prescription, the stars represent the YFS
prescription
and the solid histogram is the exact ${\cal O}(\alpha)$ distribution. The
improvement due to the inclusion of radiation coherence by means of YFS formula
is evident: the YFS-PS spectrum fits perfectly to the exact ${\cal O}(\alpha)$
one. On the right, the photon energy distributions are compared: we note that 
the result of LL description is very good by itself (except close to the 
distribution end, where anyway the statistics is very poor), as a consequence
of the good approximation of the Altarelli-Parisi splitting function to
reproduce the correct energy spectrum. Nevertheless, the YFS-PS fits better to
the exact distribution, even at its end. The same comments apply to fig.
\ref{fig:em_dist}, where variables referring to final state fermions are
plotted for the same sample of events: electron energy (on the left)
and acollinearity (on the right) distributions are shown.

\begin{figure}
\bce
\vspace{7.5cm}
\includegraphics{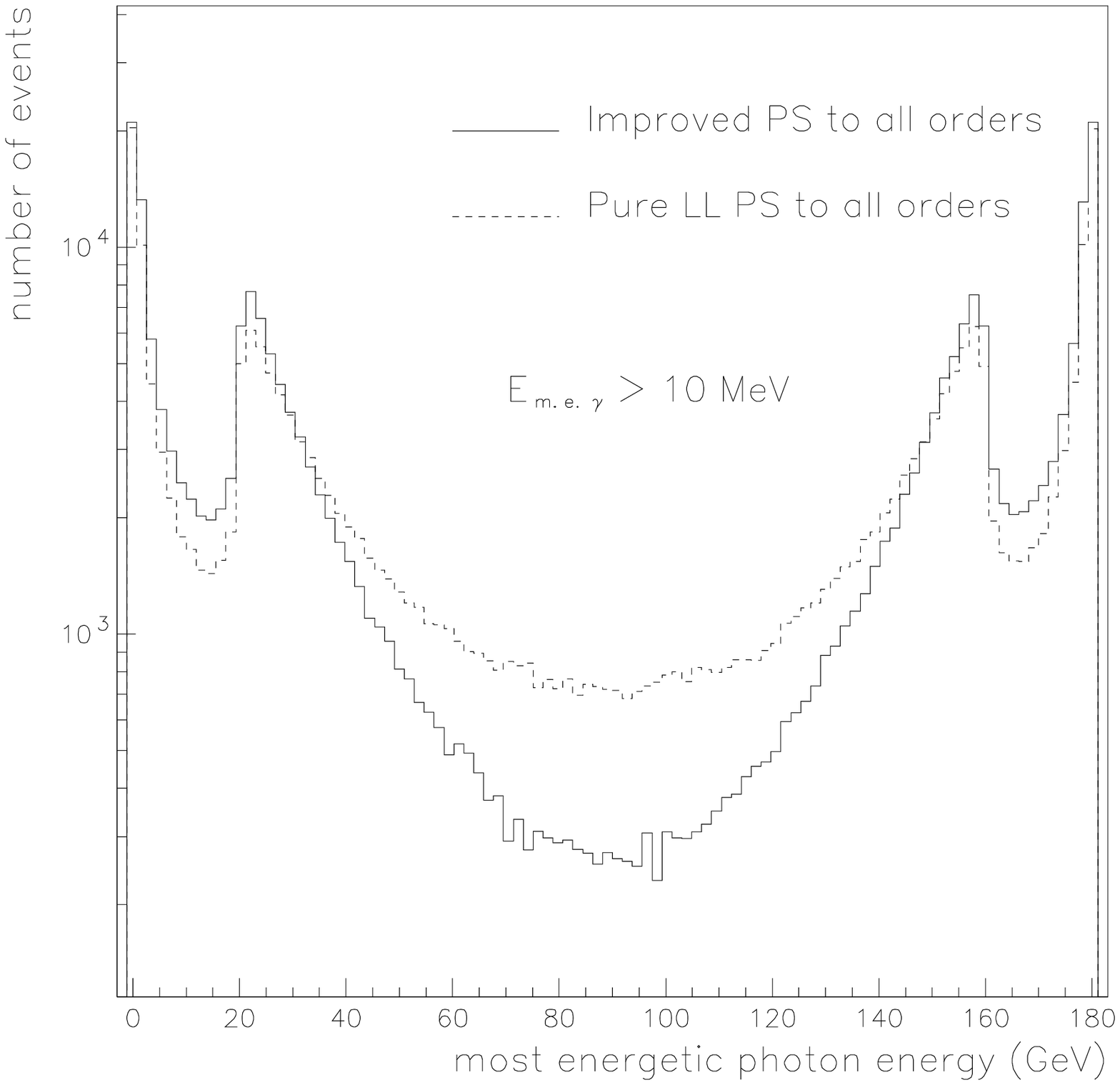}
\includegraphics{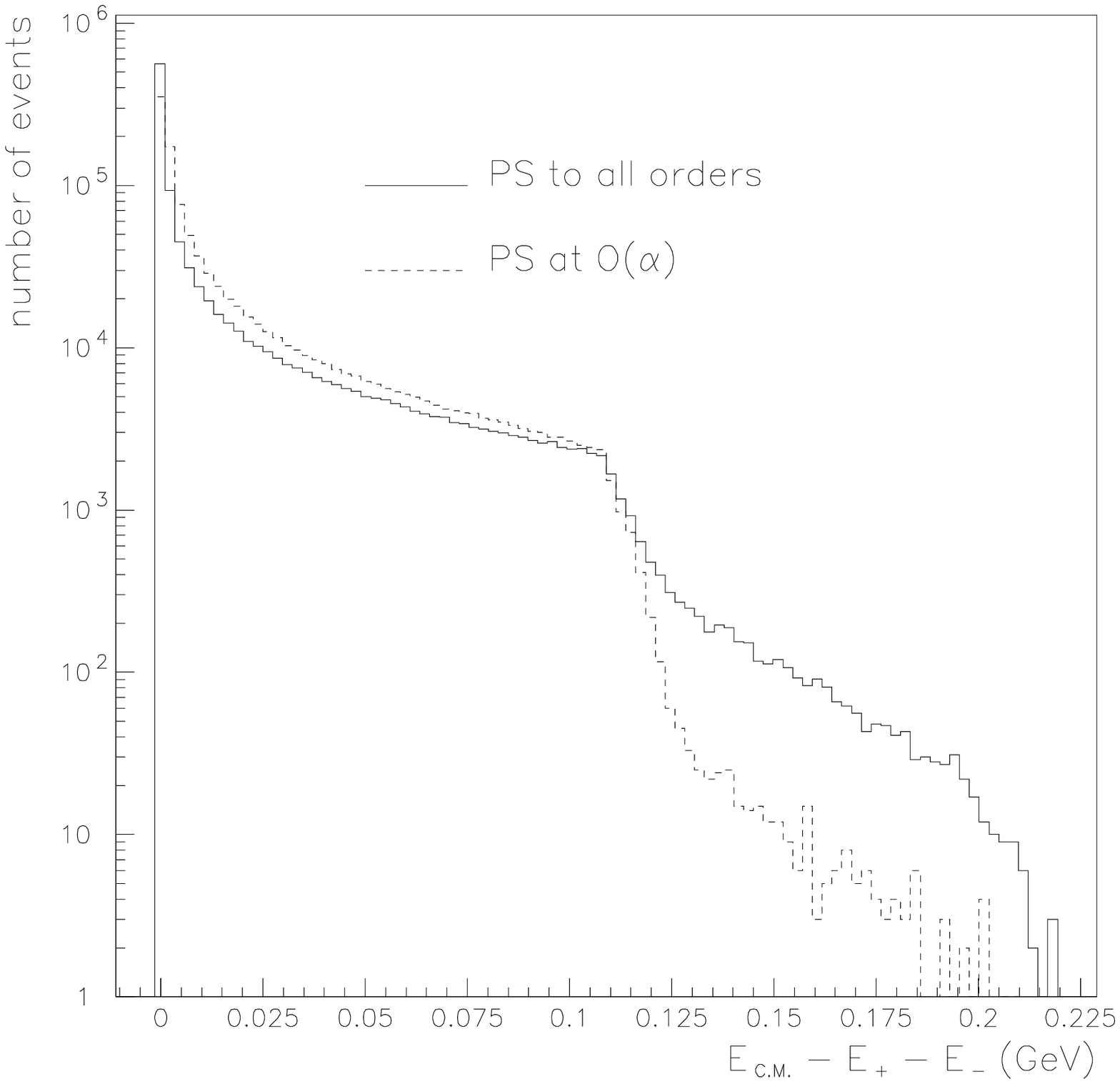}
\ece
\caption{On the left, the angular distributions of the most energetic photon as
obtained with the LL PS and YFS PS to all orders are compared for a sample of
radiative Bhabha events. On the right, the radiation energy distributions as
obtained with the ${\cal O}(\alpha)$ PS and all orders PS are compared for a
sample of inclusive Bhabha events.}
\label{allvsoalPS}
\end{figure}

Up to now, we presented the results of the ${\cal O}(\alpha)$ PS, but the YFS
prescription of eq. (\ref{cthYFS}) has been applied to the all orders PS as
well. As an example, we considered a sample of Bhabha events generated by means
of the PS to all orders, imposing the same cuts as before and requiring that the
most energetic photon of each event has at least 10 MeV of energy. The
differences (due to coherence) on the angular distribution of the most
energetic photon as given by the LL PS and the YFS PS are evident in fig.
\ref{allvsoalPS} (on the left). In the same figure, on the right, 
the effects of higher order photon emission are shown. The
distribution of the energy lost because of radiation is plotted, comparing 
the ${\cal O}(\alpha)$ and the up to all order PS distribution. In this case, no
cut is applied to the photon(s).
\begin{figure}
\bce
\vspace{7.5cm}
\includegraphics{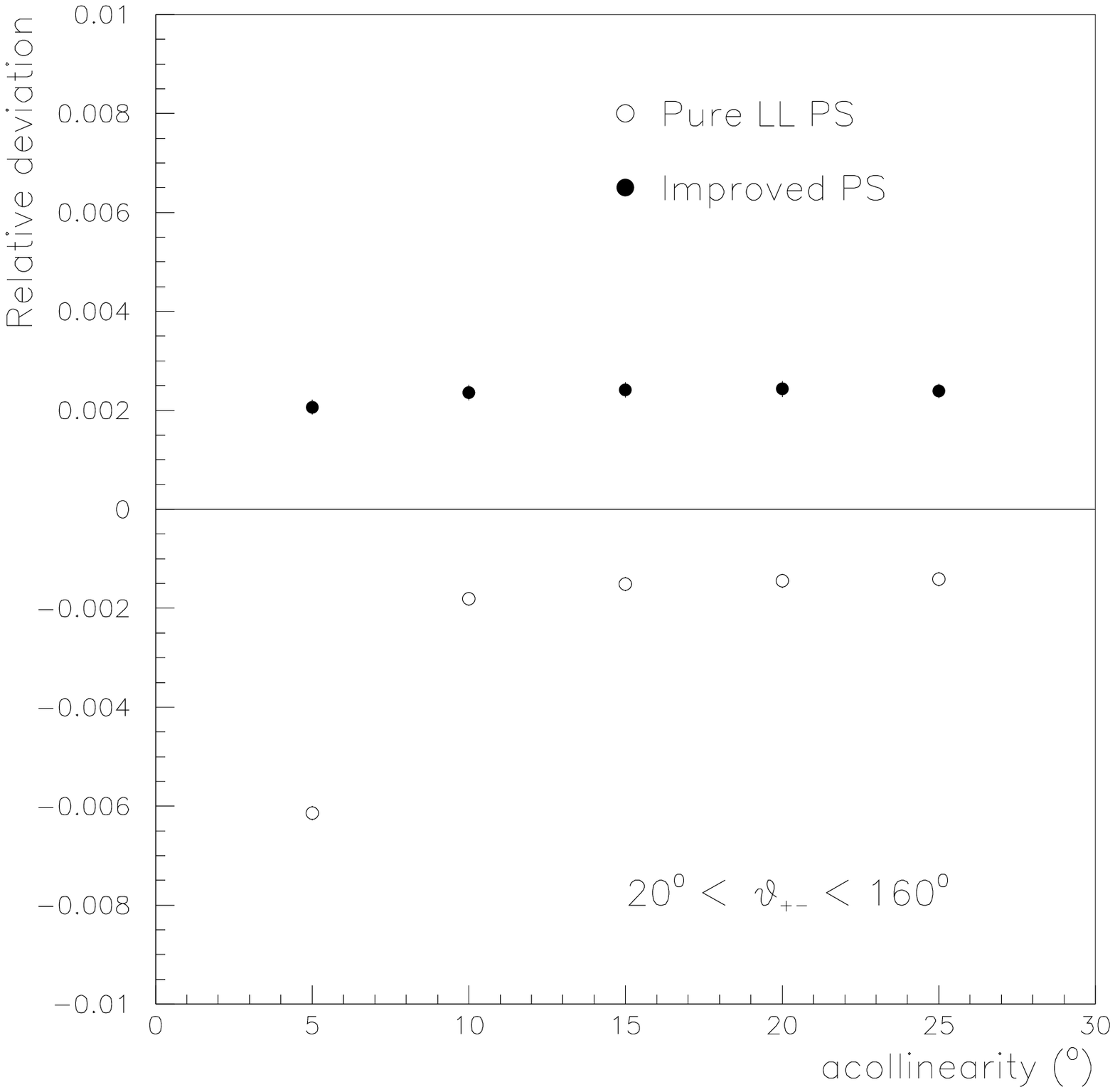}
\includegraphics{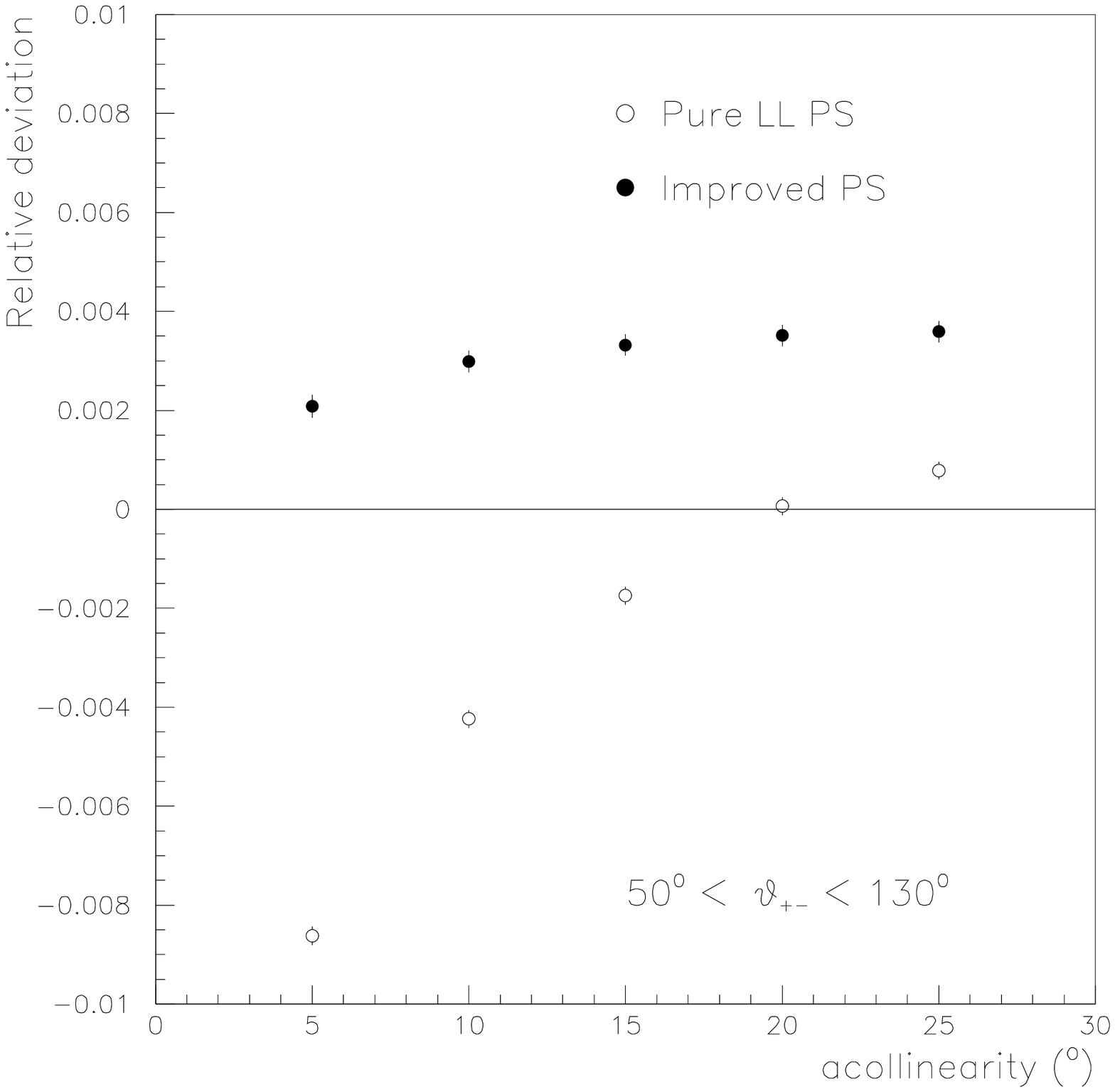}
\ece
\caption{Relative difference between ${\cal O}(\alpha)$ corrected Bhabha cross
section at $\sqrt{s}=1$ GeV, as given by the exact matrix element and the
${\cal O}(\alpha)$ PS (in the LL and YFS prescriptions). Two different angular
acceptance regions are considered and the difference is plotted as a function of
the acollinearity cut.}
\label{fig:missoa}
\end{figure}

Finally, we give an estimate of the theoretical accuracy of the PS approach in
inclusive cross section calculation. The bulk of the terms missed in the PS
scheme in its all orders implementation can be estimated, up to very
negligible uncertainty at next to leading ${\cal O}(\alpha^2)$ level
\cite{ournpb}, by means of the relative difference between
the exact ${\cal O}(\alpha)$ corrected cross section and the
${\cal O}(\alpha)$ PS one. In figure (\ref{fig:missoa}), such a difference is
plotted, as a function of the acollinearity cut, for inclusive Bhabha cross
section at $\sqrt{s}=M_\Phi$, requiring the final state $e^+$ and $e^-$ to have
at least $0.4$ GeV of energy and considering 
$20^\circ<\vartheta_\pm<160^\circ$ and $50^\circ<\vartheta_\pm<130^\circ$
acceptance region. For the ``best'' prescription, the YFS one, the amount of
${\cal O}(\alpha)$ missing contributions is contained within the 0.4\% and
they could be further reduced, by an {\em ad hoc} and tuned choice of
the $Q^2$ scale present in the electron SF. This topic is widely discussed in
\cite{ournpb}. It is worth noticing that the exact versus PS difference is less
dependent on the acollinearity cut when the YFS-improved PS is considered. The
next step toward the reduction of the theoretical error (and the complete
independence from {\em ad hoc} recipes) would be the real merging of the exact
${\cal O}(\alpha)$ matrix element into the PS. Work is in progress in this
direction.

\section{Conclusions}

The QED Parton Shower algorithm has been applied to QED processes event
generation at
$e^+e^-$ flavour factories, in order to simulate the phenomenologically
relevant radiative corrections to all orders of perturbation theory and to
provide a useful Monte Carlo event generator (BABAYAGA) for data analysis. 
The theoretical accuracy of the event generator is estimated to be within
the 0.5\% level for Bhabha scattering, in typical event selection criteria for
data analysis at flavour factories. After a brief
sketch of the theoretical framework which the PS algorithm is based on, in this
paper the photons kinematics generation is critically revised. We introduced
three different schemes for generating photons angles, named the pure PS, the LL
and the YFS prescription. In particular, the last one is inspired to the YFS
formula \cite{YFS,Muta} of eq. (\ref{YFS}). By means of an ${\cal O}(\alpha)$
PS, the results from the three prescriptions have been systematically compared
to the exact ${\cal O}(\alpha)$ ones. We pointed out and showed that the YFS-PS
has the best behaviour to simulate and generate also radiative events, fitting
very well to all the exact ${\cal O}(\alpha)$ distributions. The new release of
the BABAYAGA event generator is built on this theoretical
background. The code is available and can be downloaded from the web site {\tt
http://www.pv.infn.it/\~{}nicrosi/programs.html}.

The open issue for the next future is the merging of the exact 
${\cal O}(\alpha)$ matrix element with the PS, which is an interesting task
also for QCD simulations at high energies, as documented in
\cite{MEQCD}. The merging would make the PS independent from {\em ad hoc}
recipes and would actually shift the theoretical uncertainty to the next to
leading order $\alpha^2$ corrections.
%
%
%

\section{Acknowledgments}

My work is done in collaboration with Guido Montagna, Oreste Nicrosini and
Fulvio Piccinini, whom I wish to thank for their constant support to my work.
I'm grateful also to Achim Denig, of the KLOE Collaboration, for his
interest in our work and very useful discussions. Finally, I thank the INFN,
Sezione di Pavia, for the use of the computer facilities. The author 
acknowledges partial support from the EEC-TMR Program, Contract N.~CT98-0169.

\end{document}